\documentclass[
    ,final            
  ]
  {aipproc}

\layoutstyle{6x9}

\newcommand{\be}{\begin{equation}}
\newcommand{\ee}{\end{equation}}
\newcommand{\bea}{\begin{eqnarray}}
\newcommand{\eea}{\end{eqnarray}}
\newcommand{\bref}[1]{(\ref{#1})}
\newcommand{\nn}{\nonumber}
  
 \newcommand{\D}{\delta} 
\newcommand{\ep}{\epsilon} 
\newcommand{\T}{\theta} 
   
\newcommand{\lam}{\lambda}      
          \newcommand{\w}{\omega}
          
\newcommand{\h}{\eta} \newcommand{\W}{\Omega}
\def\CL{{\cal L}}
\def\CH{{\cal H}}
\def\CP{{\cal P}}\def\CJ{{\cal J}}\def\CO{{\cal O}}

\def\CM{{\cal M}}\def\l{{\ell}}
\def\6{\partial}\def\7{\tilde}\def\8{\hat}
\def\t{\tilde}\def\pa{\partial}
\newcommand{\Iso}{I{\hskip-0.05cm{so}}}\def\so{{so}}

\def\Deq21{{$D \hskip-1mm = \hskip-1mm 2 \hskip-1mm + \hskip-1mm1$}}
\def\Dneq21{{$D \hskip-1mm\neq\hskip-1mm 2 \hskip-1mm + \hskip-1mm1$}}

\begin{document}

\title{Deformed Maxwell Algebras and their  Realizations}



\author{Joaquim Gomis}{
  address={Departament d'Estructura i Constituents de la
Mat\`eria and ICCUB, Universitat de Barcelona, Diagonal 647, 08028 Barcelona}
}

\author{Kiyoshi Kamimura}{
  address={Department of Physics, Toho University,
Funabashi, 274-8510 Japan}
}

\author{Jerzy Lukierski}{
  address={Institute of Theoretical
 Physics, Wroclaw University, pl. Maxa Borna 9,\\ 50-204 Wroclaw, Poland}
}

\begin{abstract}
We study  all possible deformations of the Maxwell algebra. In
$D=d+1\neq 3$ dimensions there is only one-parameter deformation.
The deformed algebra is isomorphic  to $ so(d+1,1)\oplus so(d,1)$ or to
$so(d,2)\oplus so(d,1)$ depending on the signs of the deformation
parameter. We construct in the $dS (AdS)$ space a model of massive particle
interacting with Abelian vector field via
non-local Lorentz force. 
In \Deq21  the deformations depend on two parameters $b$ and $k$. We
construct a phase diagram, with two parts of the $(b,k)$ plane with
$ so(3,1)\oplus so(2,1)$ and $ so(2,2)\oplus so(2,1)$ algebras separated by
a  critical curve along which the algebra is isomorphic to
${\Iso}(2,1)\oplus so(2,1)$.
We introduce in \Deq21  the Volkov-Akulov type model for a
Abelian Goldstone-Nambu vector field described by a non-linear
action containing as its bilinear term the free  Chern-Simons
Lagrangean. 
\end{abstract}

\maketitle

\section{Introduction}\footnote{
Talk based on \cite{Gomis:2009dm} in the  XXV-th Max Born 
Symposium "Planck Scale", held in Wroclaw 29.06-3.07.2009.
, ed. R.Durka, J.Kowalski-Glikman and M.Szczochor, 
AIP Conference Proceedings Series, Melville(N.Y.).
} 

 It is known since 1970  that the lagrangian of
a massive relativistic particle in a constant electromagnetic
background $f^0_{ab}$ is invariant under a modification of Poincare
algebra that leave the background invariant, (BCR algebra
\cite{Bacry:1970ye}). Its eight generators are two Lorentz
transformations $G, G^*$,  the four space-time translations and two
central charges $Z,\tilde Z$ corresponding to the electric and
magnetic charges. These central charges appear in the commutator of
the four momenta.  Later in 1972 Schrader \cite{Schrader:1972zd},
introduced the Maxwell algebra with 16 generators, the Poincare
algebra plus six non-central extension \be
\left[P_{a},~P_{b}\right]= i~ Z_{{ab}}, \quad\quad
Z_{ba}=-Z_{{ab}}.\label{NHm1} \ee The new generators $Z_{{ab}}$
describe so called tensorial central charges\footnote{ We restrict
ourselves to Minkowski space.  Later in the literature the
tensorial central charges were introduced mostly in the Poincare
superalgebras \cite{D'Auria:1982nx} \cite{vanHolten:1982mx} \cite{de
Azcarraga:1989gm} and also in p-brane non-relativistic  Galilei and
Newton-Hooke algebras \cite{Brugues:2004an} \cite{Gomis:2005pg}
\cite{Brugues:2006yd}.}
 and satisfy the relations
\be
\left[M_{ab},~Z_{cd}\right]=-i~(\h_{bc}~Z_{ad}-\h_{bd}~Z_{ac}+\h_{ad}~
Z_{bc}-\h_{ac}~Z_{bd})
.\label{Poincare0} \ee
As we discussed above a dynamical realization of BCR algebra is
obtained by considering the
relativistic particle coupled in minimal way to the electromagnetic
{ potential} $ A_b=\frac12{f^0_{ab}}x^a$ defining the constant field
strength $F_{ab}={f^0_{ab}}$. The
second order lagrangian is \be\label{bacry2} L=-m\sqrt{-\dot
x^2}+\frac e2 {f^0_{ab}}x^a\dot x^b.\ee Note that this action is not
invariant under the whole Maxwell algebra since part of the  Lorentz
rotations is  broken by the choice of constant electromagnetic field
strength
${f^0_{ab}}$. In order to recover the Maxwell symmetry one has to
promote ${f^0_{ab}}$ to be the dynamical degrees of freedom
 and  consider an extension of space-time by supplementing the new
coordinates $\theta^{ab} (=-\theta^{ba})$ which are the group
parameters associated
to $Z_{ab}$. In order
to introduce the dynamics invariant under the Maxwell group
symmetries we have applied in \cite{Bonanos:2008kr}
\cite{Bonanos:2008ez}  the method of non-linear realizations
employing the Maurer Cartan (MC) one-forms (see e.g.
\cite{Coleman},\cite{Gomis:2006xw}).

The aim of this paper is to describe in any dimensions $D$
all possible deformations of the
Maxwell algebra \bref{NHm1}, \bref{Poincare0}, and investigate
the dynamics realizing the deformed Maxwell symmetries.
 In  \Dneq21 there exists only one-parameter deformation which
leads for positive (negative) value of the deformation parameter $k$
to an algebra that is isomorphic to the direct sum of the AdS
algebra $so(d,2)$ (dS algebra $so(d+1,1)$) and the Lorentz algebra
$so(d,1)$. This deformation for $k>0$ has been firstly obtained by
Soroka and Soroka \cite {Soroka:2004fj},\cite{Soroka:2006aj}.

 In \Deq21  one gets a two-parameter family of deformations, with
second deformation parameter $b$. The parameter space $(b,k)$ is
divided in two regions separated by the critical curve
\be A(b,k)={(\frac k3)}^3-{(\frac b2)}^2=0 \label{degencurv}\ee 
on which the deformed
algebra is non-semisimple. It appears that for $A>0 ~ (A<0)$
the deformed algebra is isomorphic to $so(2,2)\oplus so(2,1)\;
(so(3,1)\oplus so(2,1))$. On the curve \bref{degencurv} the deformed
algebra is  the direct sum of \Deq21  Poincare algebra and
\Deq21  Lorentz algebra, ${\Iso}(2,1)\oplus so(2,1)$.

In order to study the particle dynamics in the deformed cases we
consider the MC one-forms on the suitable coset of deformed Maxwell group.
Firstly we obtain, for arbitrary $D$ and $k\neq0, b=0$,
the particle model in curved and enlarged space-time $y^A=(x^a, \theta^{ab})$.
We choose the coset which leads to
the metric depending only on the space-time coordinates $x^a$.  We
derive in such a case the particle model in AdS ( for $k>0$) or dS (
for $k<0$) curved space-time with the coupling to Abelian vector
field which generalizes,
 in the theory with deformed Maxwell symmetry,
 the Lorentz force term describing the particle interaction with constant
 electromagnetic field. The Lorentz force in the case studied here
becomes non-local.

In \Deq21  and $k=0, b\neq 0$, we will consider a nonlinear field
theory realization of the deformed Maxwell algebra in six-dimensional
enlarged space $(x^a,\theta^{a}=\frac 12 \epsilon^{abc}
\theta_{bc};\,a,b=0,1,2)$ by assuming that the surface
$\theta^{a}=\theta^{a}(x)$ describes \Deq21  dimensional Goldstone
vector fields\footnote{Such a method was used firstly by Volkov and
Akulov to derive the Goldstino field action \cite{Volkov:1973ix}.}.
 If we postulate the action of Volkov-Akulov type
\cite{Volkov:1973ix} \cite{Zumino:1977av} we shall obtain the field
theory in \Deq21 space-time with a lagrangian  containing  a free
Abelian Chern-Simons term\cite{Schwarz:1978cn},
\cite{Deser:1982vy}.

 The organization of the paper is as follows. After reviewing some properties
of the Maxwell group we consider the corresponding
particle model. Further we will present all possible deformations of
Maxwell algebra. We construct also the deformed particle model for
arbitrary $D$  with $k\neq0, b=0$. For
\Deq21  case with $k=0, b\neq 0$ we promote the
group parameters $\theta^a$
to Goldstone fields  $\theta^a(x)$. These Goldstone-Nambu fields will be
described by Volkov-Akulov type action. Finally
we present a short summary.

\section{Particle Model from the Maxwell algebra (arbitrary $D$)}

A particle model invariant under the complete Maxwell algebra
can be derived geometrically \cite{Bonanos:2008ez} by
the techniques of non-linear realizations, see e.g. \cite{Coleman}.
Let us consider the coset G/H=Maxwell/Lorentz \cite{Bonanos:2008kr}
\cite{Bonanos:2008ez} parametrized by $
g=e^{iP_ax^a}e^{\frac{i}{2}Z_{ab}\theta^{ab}}.$ 
The corresponding Maurer-Cartan (MC) one-forms are \be
\label{omega}\Omega=-i g^{-1}dg= P_a~e^a+\frac12
Z_{ab}~\omega^{ab}+\frac12 M_{ab}~l^{ab}, \label{MCform}\ee
where \be e^a=dx^a, \hspace{15 pt}
\omega^{{ab}}=d\theta^{ab}+\frac12(x^a\;dx^b-x^b\;dx^a), \hspace{15
pt} l^{{ab}}=0. \label{explicit1} \ee

A second order form of the lagrangian for the particle invariant
under the full Maxwell algebra in the extended space-time $y^A=(x^a,
\theta^{ab})$  can be written using  the extra coordinates
$f_{ab}$ that transform covariantly under the Lorentz
group \cite{Bonanos:2008kr}
\be \tilde L=-m{\sqrt{-\dot x^2}}+
\frac{1}2 f_{ab}\,\left(\dot\T^{ab}+\frac12(x^a\dot x^b- x^b\dot
x^a)\right).\label{lagrangian2}\ee
{The Euler-Lagrange} equations of motion are \bea\label{eqmotion1theta}
\dot f_{ab}&=&0,\qquad 
\dot\T^{ab}=-\frac12(x^a\dot x^b-x^b\dot x^a),\qquad 
m \frac{d}{d\tau}\frac{\dot x_a}{\sqrt{-\dot
x^2}}=f_{ab}\dot x^b. \label{eqmotion1xx}\eea
Integration of \bref{eqmotion1theta} gives
$f_{ab}=f^0_{ab}$ and such a solution breaks the Lorentz symmetry
spontaneously into the BCR subalgebra of Maxwell algebra. Substituting
this solution in the $x$-equation of motion in \bref{eqmotion1xx} we
provide the motion of a particle in the constant electromagnetic field
\cite{Bacry:1970ye},\cite{Schrader:1972zd} described by the lagrangian
\bref{bacry2}.
 Notice that the interaction part of the lagrangian \bref{lagrangian2}
defines an analogue of the EM potential $\hat A$ as one-form
 in the extended bosonic space $(x,\theta, f)$
\be \hat
A=\frac{1}2\,{f}_{ab}\,\omega^{ab}, \qquad with\qquad
\hat F=d\8A=\frac{1}2\,{f}_{ab}\,e^a\wedge
e^b +\frac{1}2\,{df}_{ab}\wedge\omega^{ab}.\label{AMax}
\ee
We see that the
field strength has the constant components ${f}_{ab}$  on-shell, $\dot f_{ab}=0$ .


The infinitesimal symmetries of the lagrangian \bref{lagrangian2}
are realized canonically as
Noether 
generators 
\bea
\CP_a&=&-\left(p_a-\frac12\,p_{ab} x^b\right),\qquad
{\cal Z}_{ab}=-\,p_{ab},
\nn\\
\CM_{ab}&=&-\left(p_{[a} x_{b]}+ p_{[ac}{\T_{b]}}^c+
p^f_{[ac}{f_{b]}}^c \right),
 \label{cgenM}\eea
where
$p_a,p_{ab},p_f^{ab}$ are the canonically conjugated momenta of the
coordinates  $x^a, \theta^{ab}, f_{ab}$.
From the lagrangian \bref{lagrangian2} we obtain the constraints \bea
\phi&=& \frac{1}{2}\left(\pi_a^2+m^2\right)=0,\qquad
\phi_{ab}=p_{ab}-f_{ab}=0,\qquad
\phi_f^{ab}=p_f^{ab}=0,
\label{constraintMax}\eea
where $\pi_a\equiv p_a+\frac12\,f_{ab}\,x^b$.
The last two are the second class constraints
and are solved as 
$(f_{ab},p_f^{ab})=(p_{ab},0)$.
The Hamiltonian becomes \be \CH=\lam\,\phi=\frac{\lam
}{2}\left(\pi_a^2+m^2 \right) \ee
and
the constraints \bref{constraintMax}  and the global
generators \bref{cgenM} are shown to be conserved.

There are four Casimirs in the Maxwell algebra in four dimensions,
\cite{Schrader:1972zd},\cite{Soroka:2004fj}
\bea C_1&=&\CP_a^2-\CM_{ab}{\cal Z}^{ab},\qquad C_2=\frac12\,{\cal Z}_{ab}^2,
\nn\\
C_3&=&({\cal Z}\t {\cal Z}),\qquad 
C_4=(\CP^{b}\t {\cal Z}_{ba})^2+ \frac14({\cal Z}\7{\cal Z})\,(\CM_{ab}\,\t {\cal Z}^{ab}),
\label{4Casimir}\eea
where  $\7{\cal Z}^{ab}=\frac12\ep^{abcd}{\cal Z}_{cd}$.
In the first quantized theory they are imposed in the Schr\"odinger
representation. In particular the first one gives the
generalized KG equation,
\be\left[\left(\frac1{i}\frac{\pa}{\pa x^a}+\frac1{2i}x^b
\frac{\pa}{\pa \T^{ab}}\right)^2+m^2\right]\Psi(x^a,\T^{ab})=0. \ee

\section{Deformations of Maxwell algebra}

In this section we would like to find all possible deformations of
the Maxwell algebra. The problem of finding the continuous
deformations of a Lie algebra can be described in cohomological
terms \cite{levynahas}, we follow the notations of
ref.\cite{Gibbons:2007iu}.

The MC form for the Maxwell algebra is
\be \Omega=P_a L_P^a+\frac 12 Z_{ab}L_Z^{ab}+\frac 12 M_{ab} L_M^{ab}\ee
and the MC equations in this case are given by\footnote{As usual we will
often omit "$\wedge$" for exterior product of forms.}
  \bea
dL_M^{ab}+L_M^{ac}{L_{Mc}}^b&=&0,
\nn\\
dL_P^{a}+L_M^{ac}{L_{Mc}}&=&0,
\nn\\
dL_Z^{ab}+L_Z^{ac}{L_{Mc}}^b+L_M^{ac}{L_{Zc}}^b-L_P^aL_P^b&=&0.
\label{defM}\eea
The infinitesimal deformations are characterized
by the non-trivial vector-valued  two-forms $A^{(2)}$ verifying
 \be DA^{(2)} =0\,,\qquad A^{(2)} \ne -D\Phi^{(1)}.  \label{nontriv} \ee
The non-trivial infinitesimal deformations are  in one to one
correspondence with the second cohomology group $H^2( g;g)$.
In the case when $H^3( g; g)$ vanishes, it is always possible to
choose a representative in  the class of infinitesimal deformations
such  that it verifies the Jacobi identity in all orders.

Solving \bref{nontriv} we find a one-parameter family of non-trivial
solutions for $A^{(2)}$, with the exception that there is a
two-parameter family in "exotic" case \Deq21.\footnote{ Some of the
calculations with forms are being done using the Mathematica code
for differential forms EDC \cite{bonanos}.}
The MC equations get additional terms representing deformations as follows
\bea dL_M^{ab}+L_M^{ac}{L_{Mc}}^b&=&b\,\ep^{abc}L_{Zcd}L_P^d,
\nn\\
dL_P^{a}+L_M^{ac}{L_{Pc}}&=&k\,L_Z^{ac}{L_{Pc}}+b\,\frac14L_Z^{ab}\ep_{bcd}L_Z^{cd},
\nn\\
dL_Z^{ab}+L_Z^{ac}{L_{Mc}}^b+L_M^{ac}{L_{Zc}}^b-L_P^aL_P^b&=&k\,L_Z^{ac}{L_{Zc}}^b,
\qquad
 (\ep^{012}=-\ep_{012}=1).
\label{defQ2}\eea
Here $k$ and $b$ are arbitrary real constant
parameters; we stress that deformation terms proportional to $b$
are present only  in \Deq21.
The length dimensions of $k$ and $b$ are respectively $[L^{-2}]$ and
$[L^{-3}]$.

The general deformed Maxwell algebra found in the previous
subsection can be written in terms of the commutators of generators.
In general dimensions there exists only the following $k$-deformed algebra,
with $b=0$
\bea
\left[P_a,~P_b \right]&=& i~{Z}_{ab},\qquad\qquad
\left[M_{ab},~M_{cd}\right]=-i~\h_{b[c}~M_{ad]}+i~\h_{a[c}~M_{bd]},\nn\\
\left[P_a,~M_{bc}\right]&=&-i~\h_{a[b}~P_{c]}, \qquad
\left[Z_{ab},~M_{cd}\right]=-i\left(~\h_{b[c}~Z_{ad]}-~
\h_{a[c}~Z_{bd]}\right),
\nn\\ 
\left[P_a,~Z_{bc}\right]&=&+ik\,\h_{a[b}~P_{c]},
\qquad
\left[Z_{ab},~Z_{cd}\right]=+ik\left(~\h_{b[c}~Z_{ad]}-i~\h_{a[c}~Z_{bd]}
\right). 
\label{defMaxalg}
\eea
For $k\neq0$ case we introduce dimensionless rescaled generators as
$\CP_{a}=\frac{P_{a}}{\sqrt{|k|}}, \CM_{ab}=-\frac{Z_{ab}}{k}, 
\CJ_{ab}=M_{ab}+\frac{Z_{ab}}{k}, $ 
then the $k$-deformation of Maxwell algebra becomes
 \bea
\left[\CP_a,~\CP_b\right]&=&-i\,\frac{k}{|k|}\,~{\CM}_{ab}, \nn\\
\left[\CP_a,~\CM_{bc}\right]&=&-i~\h_{a[b}~\CP_{c]},\qquad
\left[\CM_{ab},~\CM_{cd}\right]=-i~\h_{b[c}~\CM_{ad]}+i~\h_{a[c}~\CM_{bd]},
\nn\\ 
\left[\CP_a,~\CJ_{bc}\right]&=&\left[\CM_{ab},~\CJ_{cd}\right]=0,\quad
\left[\CJ_{ab},~\CJ_{cd}\right]=-i~\h_{b[c}~\CJ_{ad]}+i~\h_{a[c}~\CJ_{bd]}.
\label{kdefalge}  \eea

The algebra of $(\CP_a,~\CM_{cd},~\CJ_{cd},)$ for $k>0$
($k^+$-deformation) is $\so(D-1,2)\oplus \so(D-1,1)$, {\it i.e.}
we obtain the direct sum of $AdS_D$ and $D$-dimensional Lorentz group. For $k<0$
($k^-$-deformation) we get $\so(D,1)\oplus \so(D-1,1)$, i.e., the direct
sum of $dS_D$ and $D$-dimensional Lorentz group. We recall here that the
above algebra for $k>0$ was previously found by Soroka and Soroka
\cite{Soroka:2006aj}. In our further discussion we will also use the
notation $k=\pm\frac1{R^2}$ where $R$ is the radius of AdS
($k>0$) or dS $(k<0)$ space.


Three dimensional case is interesting since there is an exotic $b$-deformation
of the Maxwell algebra in addition to the $k$-deformation.
For $b=0,\;k\neq 0$ , as was discussed previously,
the algebra is $\so(2,2)\oplus\so(2,1)$ for $k>0$ ($k^+$-deformation) and
$\so(3,1)\oplus\so(2,1)$ for $k<0$ ($k^-$-deformation).
For $k=0,\;b\neq 0$ ($b$-deformation)
we can introduce
\bea
\pmatrix{\CP_a \cr \CM_a \cr \CJ_a}&=&
\pmatrix{&&\cr& U_b&\cr&& }\pmatrix{P_a \cr M_a \cr Z_a},\qquad
U_b=
\pmatrix{\frac1{\sqrt3\,{b^{1/3}}}&0&\frac1{\sqrt3\,{b^{2/3}}}\cr
-\frac1{3\,{b^{1/3}}}&\frac23&\frac1{3\,{b^{2/3}}}\cr
\frac1{3\,{b^{1/3}}}&\frac13&-\frac1{3\,{b^{2/3}}}},
\label{transbdef}\eea
where $M^a=\frac12\ep^{abc}M_{bc},
 Z^a=\frac12\ep^{abc}Z_{bc}$. The algebra becomes 
\bea
\left[\CP_{a}, \CP_{b}\right]&=&-i\,\ep_{abc}\CM^{c}, \quad
\left[\CP_{a}, \CM_{b}\right]=i\ep_{abc}\CP^{c},\quad
\left[\CM_{a}, \CM_{b}\right]=i\ep_{abc}\CM^{c},
\nn\\
\left[\CJ_{a}, \CJ_{b}\right]&=&i\ep_{abc}\CJ^{c},\quad
\left[\CP_{a}, \CJ_{b}\right]=
\left[\CM_{a}, \CJ_{b}\right]=0.\qquad
\label{J1J2alg}\eea
Then $(\CP_a,\CM_a)$ are the $\so(3,1)$ generators and $\CJ_a$ describes
$\so(2,1)$. This algebra is isomorphic to the one with
$b=0, k<0$ ($k^-$-deformation).

To examine more general case with any values of the deformation
parameters $(b,k)$ we consider the Killing form $g_{ij}=C_{ik}^\l C_{\l j}^k$
of the algebra.
Its determinant is \be \det g_{ij}=6^94^3\,A(b,k)^3,\qquad
A(b,k)\equiv (\frac{k}{3})^3-\,(\frac{b}2)^2.
\label{degenerate}\ee
In the case $\det g=0$  the Killing form is degenerate,
otherwise the algebra is semisimple.
In table 1 we summarize the cases accrding to the values of  $(k,b)$.
\begin{center}
\begin{tabular}{|c|c|c|c|c|}
  \hline
  I & $\det g=0$ & $ b=0,\; k=0,$& {Maxwell} & Maxwell algebra \\
 II & $\det g=0$ & $A(b,k)=0$& Poincar\'e&
${\Iso}(2,1)\oplus\so(2,1)$ \\
III & $\det g>0$ & $A(b,k)>0$ & AdS & $\so(2,2)\oplus\so(2,1)$ \\
IV & $\det g<0$ &   $A(b,k)<0$& dS &$\so(3,1)\oplus\so(2,1)$\\
  \hline
\end{tabular}

Table 1:
{\it The phase sectors for deformed \Deq21 Maxwell algebra}
\end{center}

\section{Particle models on the ${k}$-deformed Maxwell algebra }

In this section we will discuss a model realizing in arbitrary dimension $D$
the deformed Maxwell algebras and look for the physical meaning of the
additional coordinates $(f_{ab},\T^{ab})$.
Generalizing the results described earlier for the standard Maxwell algebra
to those for the deformed Maxwell algebra with $(k\neq0,b=0)$,
we describe a  particle interacting with constant
electromagnetic field in a generalized AdS(dS) space-time.

 We consider a coset ${G}/{H}$ with $G=\{P_a,M_{ab},Z_{ab}\},
H =\{M_{ab}\}$ and parametrize the group element $g$ using
 $(x^a,\theta^{ab})$ as  
$g=e^{iP_ax^a}e^{\frac{i}{2}Z_{ab}\theta^{ab}}.$
The MC form for $\W$ this coset is
\bea
L^a_P&=&{e}^b\,{\Lambda_b}^a,\quad
L_Z^{cd}=-\frac1{k}\,{\Lambda^{-1c}}_a\,\left(\w^{ab} -({\Lambda}\,d
{\Lambda^{-1}})^{ab}\right){\Lambda_b}^d,\quad
L_M^{cd}=0, \label{MCForms}\eea
where $(e^a,\w^{ab})$ are AdS(dS) drei-bein and spin connection satisfying
\be de^a+{\w^a}_be^b=0,\qquad
d\w^{ab}+{\w^a}_c\w^{cb}=-k\,e^ae^b \ee
and  ${\Lambda}$ is a vector Lorentz transformations (Lorentz harmonics)
in terms of  new tensorial coordinates $\T^{ab}$ as follows
\be
{\Lambda_a}^b={(e^{-k\T})_a}^b={\D_a}^b+{(-k\T)_a}^b+\frac1{2!}{(-k\T)
_a}^c{(-k\T)_c}^b+...\, .\label{Lamab}\ee

 The particle action generalizing  \bref{lagrangian2} for $k\neq 0$ looks as follows
\bea
\CL\,d\tau=-m\sqrt{-\h_{ab}\,L_P^{a*}L_P^{b*}}+\frac12f_{ab}L_Z^{ab*}=
-m\sqrt{-g_{ab}(x)\,\dot x^a\dot x^b}\,d\tau+\8A^*,
\qquad  \label{Lag50}\eea
where $g_{ab}$ is the AdS(dS) metric, now depending only on $x$,
\bea
g_{ab}&=&{e_a}^c{e_b}^d\h_{cd}=
\h_{ab}+\left[(\frac{\sin(\sqrt{kr^2})}{\sqrt{kr^2}})^2-1\right]
(\h_{ab}-\frac{x_a x_b}{x^2}). \label{metric1}\eea
We obtain the metric of AdS (dS)
space with radius $R$, where $k=1/R^2$. The
pullback $\8A^*$ in \bref{Lag50} takes the explicit form
\bea
\8A^*&=&-\frac12f_{ab}L_Z^{ba*}
=-tr(\frac12f\,L_Z^{*})
=+\frac1{2k}\,tr\left[
f\,\Lambda^{-1}\left( \w_\tau-\Lambda\,\pa_\tau
\Lambda^{-1}\right)\Lambda\right]\,d\tau,
\nn\\
&& \w_\tau^{cd}=\frac{\dot x^{[c}x^{d]}}{{x^2}}\,
(\cos({\sqrt{kr^2}})-1). \label{Lag3}\eea In a limit $k\to 0$
(equivalently $R\to\infty$) we obtain the undeformed Maxwell case
\bref{AMax} \cite{Bonanos:2008ez}.

Now we shall describe the equations of motion following from the
lagrangian \bref{Lag50}. They are (for details see \cite{Gomis:2009dm})
\be
\w_\tau-\Lambda\,\pa_\tau
\Lambda^{-1}=0,\qquad
\dot f_{ab}=0,\qquad
m\,\nabla_\tau^2 x_a=F_{ab}\dot x^b,
\label{eomx0}\ee
where $\nabla_\tau^2$ is covariant second derivative. The field strength is
\bea
F_{ab}(x,\theta) 
&=&\7f_{ab}\,
\left(\frac{\sin(\sqrt{kr^2})}{\sqrt{kr^2}}\right)^2-
\frac{\7f_{[ac}x^cx_{b]}}
{x^2}\,\frac{\sin(\sqrt{kr^2})}{\sqrt{kr^2}}\,
\left(\frac{\sin(\sqrt{kr^2})}{\sqrt{kr^2}}-1\right),
\label{Fmn}\eea
where $ \7f_{ab}
=(\Lambda\,f\Lambda^{-1})_{ab}.$ We see that for $k\neq0$ the
generalized Lorentz force depends on $\T^{cd}$ but  in the limit
$k\to 0$ we get $F_{ab}=f_{ab}$ as expected.  If we use
\bref{eomx0} to express $\Lambda$ in the terms of the 
coordinates $x^a$, we obtain a non-local Lorentz force.

\section{$b$-deformed Maxwell algebra in D=2+1  and 
Goldstone-Nambu vector fields}

Let us consider now the deformation $k=0,b\neq0$ in \Deq21.
As an application of the $b$-deformed Maxwell algebra we discuss
the Volkov-Akulov formula\cite{Volkov:1973vd},\cite{Ivanov:1975zq}
for invariant \Deq21 Goldstone field action,
\be
S=\int\,(-\frac1{3!})\ep_{abc}\,L_P^{a\star}L_P^{b\star}L_P^{c\star}=
\int\,d^3 x\, \CL_\theta,\qquad \CL_\theta= \det({\tilde e^a}{}_b),
\label{vaaction}\ee
where $L_P^{a\star}$ is the pullback with respect to $x^a\to\T^a$, then
$ d\theta^{a\star}=\frac{\partial\theta^{a}(x)}{\partial
x^b} dx^b ; $ in such a way the Lagrangian density is a function of
$\T(x)$ and $\pa_a\T(x)$.
The MC forms are obtained as in the previous cases.
Using the detailed expressions in \cite{Gomis:2009dm} we obtain, 
for the small $b$,
\bea \CL_\theta&=&\det ({\tilde e^a}{}_b)=1-b\left(
(x\T)+\frac12\ep_{abc}\T^a\frac{\partial\theta^{c}}{\partial x_b}\right)
+b^2\left(
(-\frac{3(x^2)^3}{2240}+\frac{x^2\,\T^2}{12}+\frac{(x\T)^2}3)\right.
\nn\\&-&\left.
\frac{x^2}{48}((x\T){\D_j}^i-x_j\T^i)\frac{\partial\theta^{j}}{\partial x^i}
+\frac{(x\T)}4\ep_{abc}\T^a\frac{\partial\theta^{c}}
{\partial x_b}+
\frac18\,\ep_{abc}\ep^{def}\T^a\T_d \frac{\partial\theta^{b}}{\partial x^e}\frac{\partial\theta^{c}}
{\partial x^f}\right)+\CO(b^3).
\nn\\ \label{VAlag}\eea
The lagrangian density \bref{VAlag} contains as one of two terms linear
in $b$ the exact topological lagrangian for \Deq21 Chern-Simons field
\bea \CL^{CS}&=&-\frac{b}2\,{\epsilon_{abc}}\theta^a \frac{\partial\theta^{b}}
{\partial x_c}.
\eea
If we consider higher order terms in $b$ they can be treated
as describing new interaction vertices implying that
the Nambu-Goldstone field $\T^a(x)$ looses its topological nature.
The appearance of the terms depending explicitly
on $x^a$ and $\T^a$ in \bref{VAlag} is
related with the curved geometry in the extended space $(x^a,\T^a)$.

\section{Outlook}

In this paper  we have reviewed the deformations of the Maxwell
algebra studied in \cite{Gomis:2009dm}. The general mathematical
techniques permit us to solve the problem of complete classification
of these deformations. The commuting generators $Z_{ab}$ in
\bref{NHm1} become non-abelian in arbitrary dimension $D$ and they
are promoted to the $\frac{D(D-1)}{2} $ generators of $so(D-1,1)$
Lorentz algebra. The particle dynamics in the $\frac{D(D+1)}{2} $
dimensional coset with generators $(P_a,Z_{ab})$ becomes the theory
of point particles moving on AdS (for $k>0$) or dS   (for $k<0$)
group manifolds in external electromagnetic fields. If we use
standard formula \bref{Lag50} for the particle action in curved
space-time one can show that the particle moves only in the
space-time sector $(x^a,\T^{ab}=0)$ of the extended  space-time
$(x^a,\T^{ab})$ with a non-local Lorentz force.

In "exotic" dimension \Deq21  the symmetry corresponding to the two parameter
deformation of  Maxwell algebra is less transparent.
The coset with generators $(P_a,Z^{a})$ and coordinates $(x^a,\T_{a})$
in \Deq21
if $b\neq0$ is neither the group manifold nor even the symmetric coset space.
In order to find the dynamical realization we
assume that the coordinates $x^a$ are primary and the coordinates
$\T^a$ describe the Goldstone field values. We derive a non-linear
lagrangian for vector Goldstone field containing the bi-linear kinetic
term describing exactly the  \Deq21 CS Abelian action.

{ Finally we would like to point out some issues which
deserve further investigation:}

1) It is interesting to consider the supersymmetric extension of the BCR
algebra,
describing the symmetries of the space-time in the presence of
constant electromagnetic field background,
to the supersymmetries of superspace in general backgrounds of SUSY gauge
fields. Further question is the formulation of the SUSY extension of
the Maxwell algebra. These issues are under
consideration\cite{Bonanos09}.

2) As we already mentioned, the deformation parameter $k$ with the
dimensionality $[L^{-2}]$ can be described by the formula $|k|=\frac1{R^2}$,
and interpreted as the AdS(dS) radius for $k>0(k<0)$. The  parameter $b$,
with the dimensionality
$[L^{-3}]$, if $k=0$ is related with the closure of the quadrilinear
relation for the following non Abelian translation generators
$P_a$,
 \be
[[P_a,P_b],[P_c,P_d]]=ib\,(\h_{a[c}\ep_{bd]e}-\h_{b[c}\ep_{ad]e})
P^e. \label{PPgeneric1k}\ee
{This relation is an example of higher order
Lie algebra for $n=4$ \cite{Hanlon:1995,DeAzcarraga:1996ts}.
It is an interesting task to understand the
translations  \bref{PPgeneric1k} as describing some \Deq21 dimensional
curved manifold.}

  3) Recently in \cite{Bonanos:2008kr}\cite{Bonanos:2008ez} they
were considered  infinite sequential extensions of the Maxwell
algebra with additional tensorial generators. The concrete form of
these extensions can be determined by studying the
Chevalley-Eilenberg cohomologies at degree two.
The point particle models related with these Poincare algebra
extensions have been studied in \cite{Bonanos:2008kr}.
There appears an interesting question of the dynamical and physical
interpretation of the additional tensorial degrees of
freedom.

\begin{theacknowledgments}

We thank Jorge Alfaro, Sotirios Bonanos, Roberto Casalbuoni, Jaume
Garriga, Gary Gibbons, Christopher Pope,  Dimitri Sorokin and
Mikhail Vasilev for discussions.
JL would like to thank Universitat de Barcelona for warm hospitality and
acknowledge the support by Polish Ministry of Science and
High Education grant NN202 318534. This work has been partially
supported by MCYT FPA 2007-66665, CIRIT GC 2005SGR-00564, Spanish
Consolider-Ingenio 2010 Programme CPAN (CSD2007-00042).
J.G. would like to
thank the Galileo Galilei Institute for Theoretical Physics for its
hospitality and INFN for partial support during part of the
elaboration of this work.

\end{theacknowledgments}

\end{document}